\documentclass[twocolumn]{autart}    % Enable this line and disable the 
\usepackage{xcolor, graphicx}          % Include this line if your 
\usepackage[hyphens]{url}
\usepackage{amsmath, amsfonts, mathtools, amssymb, bm, commath}

\DeclareMathOperator{\trace}{\operatorname{tr}}
\DeclareMathOperator{\expect}{\mathbb{E}}
\DeclareMathOperator{\probability}{\mathbb{P}}

\newcommand{\frob}{\ensuremath{\text{F}}}

\newcommand{\inv}{^{-1}}

\newcommand{\todo}[1]{{\color{red}\footnote{\color{red}#1}}}

\usepackage{etoolbox, comment, mdframed}
\newtoggle{preprint}
\settoggle{preprint}{true}
\iftoggle{preprint}{%
}{\excludecomment{skippedproof}}

\begin{document}

\begin{frontmatter}

\title{Debiasing Continuous-time Nonlinear Autoregressions\thanksref{footnoteinfo}} % Title, preferably not more 

\thanks[footnoteinfo]{This paper was not presented at any IFAC 
meeting. Corresponding author Simon Kuang.}

\author[ucd]{Simon Kuang}\ead{slku@ucdavis.edu},    % Add the 
\author[ucd]{Xinfan Lin}\ead{lxflin@ucdavis.edu},               % e-mail address 
% \author[Baiae]{Publius Maro Vergilius}\ead{vergilius@culture.ir}  % (ead) as shown

\address[ucd]{Department of Mechanical and Aerospace Engineering, University of California, Davis; Davis, California, USA}  % Please supply                                              
% \address[Rome]{Senate House, Rome}             % full addresses
% \address[Baiae]{The White House, Baiae}        % here.

\begin{keyword}                           % Five to ten keywords,  
system identification% Cicero; Catiline; orations.               % chosen from the IFAC 
\end{keyword}                             % keyword list or with the 
                                          % help of the Automatica 
                                          % keyword wizard

\begin{abstract}
We study how to identify a class of continuous-time nonlinear systems defined by an ordinary differential equation affine in the unknown parameter.
We define a notion of asymptotic consistency as \((n, h) \to (\infty, 0)\), and
we achieve it using a family of direct methods where the first step is differentiating a noisy time series and the second step is a plug-in linear estimator.
The first step, differentiation, is a signal processing adaptation of the nonparametric statistical technique of local polynomial regression.
The second step, generalized linear regression, can be consistent using a least squares estimator, but we demonstrate two novel bias corrections that improve the accuracy for finite \(h\).
These methods significantly broaden the class of continuous-time systems that can be consistently estimated by direct methods.
\end{abstract}

\end{frontmatter}

\section{Introduction}
% Identifying continuous-time laws of motion from sample data is a basic concern in engineering.
The monumental achievements of a century of control theory---filtering, smoothing, prediction, and control---depend on accurate models of dynamic systems.
Today such models can be identified inductively from an inexhaustible surplus of data.

Our paper studies the inverse problem of identifying a continuous-time nonlinear dynamic system from noisy, discrete data.
Motivation for continuous-time modeling, which we do not repeat, is found in \cite[Chapter 1]{gonzalez_continuous-time_2022} and \cite[Chapter 1]{garnier_identification_2008}.
``It is a fact that the economy does not cease to exist in between observations'' \cite{phillips_et_1988}.
Discrete data is an indelible legacy of the civilian digital revoluation: both the deliberative and reactive aspects of control run on a clocked computer.

We consider only nonlinear autoregressions: models that define a linear relationship between the highest time derivative of the measured output and nonlinear functions of the lower-order time derivatives of the measured output as well as time-varying quantities such as an exogeneous input.
An intuitive solution is to approximate derivatives from the measured time series and then to minimize the integrated squared  instantaneous model residual across time.
For linear systems, this amounts to the State Variable Filtering (SVF) method \cite[\S1.5.1]{garnier_identification_2008}.
SVF is asymptotically biased as a result of measurement noise.
This issue has spawned an industry of workarounds, including bias compensation \cite[\S7.1]{soderstrom_errors--variables_2018} and iterative instrumental variables \cite{pan_consistency_2020,pan_corrigendum_2022}.

Algorithms such as SVF, which do not require solving for roots of nonlinear equations or minima of non-convex functions, are called \emph{plug-in estimators} in statistics and \emph{direct methods} in system identification.
(We use the terms interchangeably.)
It is tempting to dismiss direct methods for their inferior statistical efficiency compared to maximum likelihood estimation \cite[Chapter 5]{vaart_asymptotic_1998}, which in our system identification problem would take the form of a state-space prediction error method \cite{ljung_system_1999}.

To the contrary, direct methods are having a renaissance due to their interpretation as instantaneous linear regression, e.g.~with sparsity as SINDy \cite{brunton_discovering_2016}, and connection to Koopman operator theory.
They are amenable to feature selection methods and online estimation, are time- and memory-efficient, and are insensitive to algorithm initialization.
It has been observed that noise degrades identification accuracy \cite[Fig.~6]{brunton_discovering_2016}, and if the data is noisy, an initial smoothing pass on the state and/or derivatives improves the regression.
Continuous-time SINDy \cite{brunton_discovering_2016} minimizes a total variation penalty, a recent work \cite{hsin_symbolic_2024} uses Gaussian process regression in time, and \cite{wentz_derivative-based_2023} uses a Picard iteration of the dynamics.
A comprehensive menu of signal processing choices for inverse problems can be found in \cite{van_breugel_numerical_2020}.
This idea---pre-smoothing the noisy data ahead of regression---is represented by our Least Squares (LS) estimator.

LS is used as a baseline in our paper.
As a direct method in the SVF family, it suffers asymptotic bias in the presence of measurement noise.
This phenomenon appreciably deteriorates estimation accuracy, and has largely been overlooked and under-theorized in the direct methods revival.
We thus propose two ways to mitigate bias.
The first, Bias Corrected (BC) is based on a convexity compensation technique generalizing \cite[\S7.1]{soderstrom_errors--variables_2018}.
The second, Instrumental Variables (IV), generates instruments by a clever signal processing of the data.
Thus we sidestep the commutative algebraic properties of linear time-invariant (LTI) operators (employed in LTI identification e.g.~the proofs in \cite{pan_consistency_2020}), which are not suitable for nonlinear systems.

A non-exhaustive survey of system identification work that shares one or more characteristics with our problem is found in Table~\ref{table:methods}.\footnote{For a more comprehensive bibliography, see \cite[Chapter 1]{garnier_identification_2008}.}
In Appendix \ref{section:related}, we engage more heartily with the related literature, and ultimately conclude that there is no consistency proof for a direct method for estimating continuous-time nonlinear systems from discrete data, measured with noise.

\begin{table*}
  \centering
  \begin{tabular}{lllllll}
    Method                          & (in)direct & CT/DT      & (non)linear & noise       & (non)-asymptotic & consistent?
    \\\hline
    Our OLS (\S\ref{section:ols})   & direct     & continuous & nonlinear   & measurement & asymptotic       & Thm.~\ref{thm:ols-consistency}
    \\
    Our BCLS (\S\ref{section:bc1})  & direct     & continuous & nonlinear   & measurement & asymptotic       & Thm.~\ref{thm:bcls-consistency}
    \\
    % Our BCLS2 (\S\ref{section:bc2}) & direct     & continuous & nonlinear   & measurement & n/a              & unknown
    % \\q
    Our IV (\S\ref{section:iv})     & direct     & continuous & nonlinear   & measurement & asymptotic       & Thm.~\ref{thm:iv-consistency}
    \\\hline
    PEM                             & indirect   & any        & nonlinear   & any         & asymptotic       & Yes \cite{ljung_system_1999}
    \\
    Least squares                   & direct     & discrete   & linear      & process     & asymptotic       & Yes \cite{ljung_system_1999}
    \\
    Least squares                   & direct     & discrete   & linear      & process     & non-asymptotic   & Yes \cite{ziemann_tutorial_2023}
    \\
    SVF
                                    & direct     & continuous & linear      & measurement & n/a              & No \cite{garnier_identification_2008}
    \\
    SVF                             & direct     & continuous & nonlinear   & measurement & n/a              & unknown \cite{niethammer_parameter_2001}
    \\
    modulating function             & direct     & continuous & nonlinear   & measurement & n/a              & unknown \cite{niethammer_parameter_2001,unbehauen_identification_1997}
    \\
    SRIV                            & direct*    & continuous & linear      & measurement & asymptotic       & Local \cite{pan_consistency_2020,pan_corrigendum_2022}
    \\
    finite diff.
                                    & direct     & continuous & linear      & process     & asymptotic       &
    % \((n, h) \to (\infty, 0)2\)
    Yes \cite{fan_estimation_1999,dinh-tuan_pham_estimation_2000,soderstrom_least_1997,soderstrom_bias-compensation_1997}
    \\
    BCLS                            & direct     & discrete   & nonlinear   & measurement & asymptotic       & Yes \cite{piga_bias-corrected_2014}
    \\
  \end{tabular}
  \caption{\label{table:methods}%
  Comparison of selected system identification methods.
    Further discussion in \S\ref{section:related}.
    The ``consistency?'' column assumes that persistency of excitation and other system conditions are met. We omit convergence rates for comparison methods, as the problem formulation and/or result preclude apples-to-apples comparisons. *SRIV is a fixed-point iteration whose fixed point is asymptotically consistent, but is not guaranteed to converge for all datasets.}
\end{table*}

\section{Notation}
If \((X_n)\) is a sequence in a Banach space and \((a_n)\) is a sequence of positive numbers, the asymptotic notation \(X_n = O(a_n)\) or \(X_n \lesssim a_n\) means that \(\limsup_{n} X_n/a_n < \infty\).
If \((X_n)\) is a sequence of Banach-valued random variables and \((a_n)\) is a sequence of positive numbers, the stochastic asymptotic notation \(X_n = O_p(a_n)\) means that for every \(\epsilon > 0\), there exists some \(M > 0\) such that \(\limsup_n \probability\del{a_n^{-1} \left\|X_n\right\| > M} < \epsilon\).

The notation \([a\ldots b]\) refers to the set of integers between \(a\) and \(b\) (inclusive).

We write \(\partial_k \phi\) to denote the partial derivative with respect to the \(k\)th argument of \(\phi\).

The variable \(x\) will often refer to the vector-valued state variable, and may carry up two superscripts and a subscript:
\(x^{\ell, (d)}_j\) is the \(d\)th time derivative of the \(\ell\)th component of \(x\), evaluated at some time \(t_j\).
If \(\ell\) is omitted, then \(x^{(d)}_j\) refers to the entire vector.
% The upper index notation is  deliberately overloaded:
% for \(d \in [0\ldots m]\), \(x^d\) is the \(d\)th time derivative of a state variable \(x\) taking values in \(\mathbb{R}\);
% its estimate \(\hat x^d\) is the \(d\)th component of a vector in \(\mathbb{R}^{m + 1}\).

\section{Problem statement}
We consider a bi-infinite sequence of estimation problems parameterized by \(n\), the number of observations, and \(h\), the step size.
The induced experiment duration is \(T = nh\).
% Our estimator works along the asymptotic \(n\to \infty

Let \(x:[0, T] \to \mathbb{R}^{\mathsf d_x}\) be \(m\) times differentiable and satisfy the dynamics
\begin{align}
  \partial_t^m x(t)
   & = \phi(\partial_t^0 x(t), \ldots, \partial_t^{m-1} x(t), t)^\intercal \theta_0,
  \label{eq:model-equation}
\end{align}
where \(m\) is a positive number,
\(\theta_0 \in \mathbb{R}^{\mathsf{d}_\phi \times \mathsf d_x}\) is the true parameter, and \(\phi\) is a smooth function taking values in \(\mathbb{R}^{\mathsf d_\phi}\).

The dataset \(Z \in \mathbb{R}^n\) consists of the noisy measurements
\begin{align}
  z_i & = x(ih) + \epsilon_i, \quad i \in [1 \ldots n],
  \label{eq:observation-model}
\end{align}
where \(\{\epsilon_i\}_{i \in [1 \ldots n]}\) are independent random variables satisfying \(\expect \epsilon_i = 0\) and \(\expect \epsilon_i^4 < \infty\) independent of \(n, h\).
We assume that \(\expect \epsilon_i \epsilon_i^\intercal = \Sigma_\epsilon\) is known or reliably estimated.

The following two assumptions are necessary for discretizing \eqref{eq:model-equation}.
\begin{assum}[Space regularity of \(\phi\)]
  For all \(t \in [0, T]\), all mixed first through third derivatives of \(\phi(\cdot, t)\) are bounded independent of  \(n, h\).
\end{assum}

\begin{assum}[Time regularity of \(x\)]
  \label{assum:smoothness}
  There exists a \(p > m\) such that
  for all \(k \in [0\ldots p]\), we have
  \begin{align*}
    R_{k}
     & := \sup_{t \in [0, T]} \left|\partial_t^k x(t)\right| < \infty
  \end{align*}
  independent of \(n, h\).
\end{assum}

Informally, we seek:
\begin{prob}
  Find an estimator \(\hat \theta\), given as a function of \(Z\), such that for large \(n\) and small \(h\), \(\hat \theta\) is asymptotically close to \(\theta_0\).
\end{prob}

\section{Solution idea}
The methods we present in this paper are variations on a theme:
estimate \(\theta_0\) by treating \eqref{eq:model-equation} as a linear regression with measurement error.

\subsection{Regression specification}
Let \(\{t_j\}_{j \in [1\ldots n']} \subset [0, T]\) be a set of times, numbering \(n'\) in total, used to evaluate a regression.

The observed form of \eqref{eq:model-equation} is\footnote{Following a notation convention in system identification \cite{soderstrom_errors--variables_2018} with the main difference being that \(Y\) is an unobserved higher-order derivative of the physical state.}
\begin{subequations}
  \label{eq:regression}
  \begin{align}
    Y      & = \Phi \theta_0,
    \intertext{in other words, a regression of response (predictor) \(Y \in \mathbb{R}^{n' \times \mathsf d_x}\) onto covariates (regressors) \(\Phi \in \mathbb{R}^{n' \times \mathsf{d}_\phi}\), where
    for \(j \in [1\ldots n']\), the rows of \(\Phi\) and \(Y\) are}
    \phi_j & = \phi(x^{(0)}_j, \ldots, x^{(m - 1)}_j, t_j)
    \intertext{and}
    y_j    & = x^m_j,
  \end{align}
  and the subscript \(j\) on \(x\) denotes evaluation at \(t_j\).
\end{subequations}

\begin{assum}[Persistency of excitation]
  \label{assum:persistency}
  The
  filter times \(\{t_j\}_{j \in [1\ldots n']}\) are chosen so that
  \begin{align*}
    \liminf {n'}^{-1/2} \sigma_\text{min}(\Phi) > 0
  \end{align*}
  independent of \(n, h\).
\end{assum}

\subsection{Estimating \(x_j^d\)}
The regression \eqref{eq:regression} involves the true values of the state variables and their derivatives,
but our dataset \eqref{eq:observation-model} provides only noisy measurements of \(x^0_j\).
At the regression times \(t_j\), we generate a smoothed estimate \(\hat x_j \in \mathbb{R}^{m + 1 \times \mathsf d_x}\),
\(\hat x_j^{\ell, (d)} \approx x_j^{\ell, (d)}\).

Instantiating \eqref{eq:regression} with estimated quantities,
\begin{subequations}
  \label{eq:observed-regression}
  \begin{align}
    \hat Y      & \approx \hat \Phi \theta_0,
    \\
    \hat \phi_j & = \phi(\hat x^{(0)}_j, \ldots, \hat x^{(m - 1)}_j, t_j)
    \\
    \hat y_j         & = \hat x^{(m)}_j
  \end{align}

\end{subequations}

The smoothed derivatives \(\hat x_j\) are estimated using a linear filter, where the coefficients may be taken from local polynomial regression (\S\ref{section:accurate-time}).

\subsection{Estimating \(\theta\): three ways}
The simplest way to recover \(\theta_0\) from \eqref{eq:observed-regression} is by least squares, given by the normal equations 
\begin{align*}
  \hat \Phi^\intercal \hat Y
                           & = \hat \Phi^\intercal \hat \Phi \hat\theta_{\text{LS}}.
  \intertext{This estimator is asymptotically consistent (\S\ref{section:ols}), but can be biased due to nonlinearity.
    In particular, the gram matrix \(\hat \Phi^\intercal \hat \Phi\) is convex in \(\Phi\), which is asymptotically linear in noise \(\{\epsilon_i\}\).
    It therefore incurs a positive bias proportional to \(\sigma^2\), which ultimately leads to a downward bias in \(\hat\theta_{\text{LS}}\) \cite{soderstrom_errors--variables_2018,schennach_recent_2016}.
    One solution is to estimate and subtract this bias from the OLS normal equations, resulting in what we term the BC estimator (\S\ref{section:bc1}).
  }
  \sbr{
    \hat \Phi^\intercal \hat Y  - \hat \Sigma_{\phi y}
  }
                           & = \sbr{
    \hat \Phi^\intercal \hat \Phi - \hat \Sigma_{\phi\phi}
  } \hat\theta_{\text{BC}}.
  % % removed bcls2
  % \intertext{Another approach to bias correction is to regard \(\hat\theta_\text{OLS}\) as a black-box function of data \(Z\). From this point of view, regression dilution is attributed to the fact that \(\hat\theta_\text{OLS}\) is partly concave as a function of \(Z\). If we estimate the quadratic component (using automatic differentiation) and then correct for it, we obtain the BCLS2 estimator (\S\ref{section:bc2}).}
  % \hat \theta_\text{BCLS2} & =
  % \hat\theta_{\text{LS}} -
  % \frac{\sigma^2}{2}
  % \sum_{i= 1}^{n} \partial_{z_i}^2 \hat \theta_\text{LS},
  \intertext{A second approach to bias correction is to alter the OLS normal equations to:}
  \hat \Psi^\intercal \hat Y
                           & \approx \hat \Psi^\intercal \hat \Phi \hat\theta_{\text{IV}},
\end{align*}
where \(\hat \Psi\) is an independent approximation of \(\hat \Phi\).
% \todo{remove bcls2}
This independence means that \(\hat \Psi^\intercal \hat \Phi\) no longer incurs the leading-order bias that we attempted to correct in BCLS.
After some technical refinement, we get the instrumental variables estimator (\S\ref{section:iv}).

\subsection{Contributions}
% The basic premise of our direct method---filtering followed by regression---was first articulated the 1960s as the State Variable Filtering method for linear systems \cite{garnier_identification_2008,pan_consistency_2022}.
% Our work generalizes this method to nonlinear systems, wh
% While analogous methods for nonlinear systems exist in folklore and industrial practice, they have hitherto not been subjected to theoretical scrutiny.

Our work offers consistency proofs and attribution of the principal sources of error in these methods.
We provide analysis of design parameters (smoothing bandwidth, differentiation accuracy) and novel bias correction methods.

Mathematically, our theory decouples into two parts (regression and filtering) linked by Def.~\ref{def:filtering}, which states the fourth moment estimates for derivative estimation.

\textbf{Def.~\ref{def:filtering} is taken as a hypothesis} in the regression part of the paper, which presents a least squares estimator (\S\ref{section:ols}), raises the question of bias, and presents two solutions: a bias correction based on second moment compensation (\S\ref{section:bc1})
% % removed bcls2
% another bias correction based on second-order sensitivity (\S\ref{section:bc2}),
 and a novel instrumental variables method for continuous problems (\S\ref{section:iv}).
The error estimates in the consistency proofs refer to constants \(\alpha\), \(\beta\), and \(\gamma\).

\textbf{Def.~\ref{def:filtering} is reached as a conclusion} in the filtering part of the paper (\S\ref{section:filtering}, Appendix~\ref{section:filtering-details}), which applies local polynomial theory for nonparametric regression \cite{fan_local_2003}.
There we give formulas for achieving \(\alpha\), \(\beta\), and \(\gamma\).

\section{Statistical primitives}
\begin{defn}
  \label{def:filtering}
  Given \(n, h\), a \textbf{filter scheme} chooses
  \(\alpha > 0\),
  bandwidth \(N = h^{-\alpha}\);
  \((m + 1)\times N\) coefficient matrix \(D^d_{k}\), \(d \in [0\ldots m]\), \(k \in [1\ldots N]\);
  and defines at filter times \(\{t_j\}_{j \in [1\ldots n']}\);
  the filter outputs
  \begin{align*}
    \hat x^{\ell,(d)}_j
       & :=
    \sum_{k = 1}^N
    D^d_{k} z^\ell_{n + j + k - 2}, \ d \in [0\ldots m], \ j \in [ n'], \ \ell \in \mathsf d_x
    \\
    n' & = n - N + 2.
  \end{align*}
  This filter scheme is \((\beta, \gamma)\)-\textbf{consistent} if there exist \(\beta, \gamma > 0\) such that
  \begin{align}
    \expect \hat x_j - x_j                                           & = O(h^{\beta})
    \text{ and }
    \tag{bias}
    \\
    \del{\expect \left\|\hat x_j - \expect \hat x_j\right\|^4}^{1/4} & = O(h^{\gamma}).
    \tag{fluctuation}
  \end{align}
\end{defn}
% We informally seek guarantees for ``small \(h\), large \(n\).''
% To make this rigorous, we specify the asymptotic rate 
% at which \(h\) becomes small at the same time that \(n\) becomes large.
% at which \((h, n) \to (0, \infty)\).
% \begin{defn}[\((\alpha, \beta, \gamma)\)-consistent filtering]
% % todo fix?
%   \label{def:consistent-filter}
%   Let \(\alpha \in (0, 1)\), \(\beta, \gamma > 0\), and let \((n, h) \to (0, \infty)\).
%   A sequence of overlapping filtering schemes
%   specified by \(n'\), \(N\), \(D^d_{k}(N)\), \(\{n_{\ell, j}\}_{j \in [1\ldots n'(\ell)]}\), and \(\{t_{\ell, j}\}_{j \in [1\ldots n'(\ell)]}\), is \textbf{\((\alpha, \beta, \gamma)\)-consistent} relative to \((n_\ell, h_\ell)\) if  \begin{enumerate}
%     \item the window length \(N_\ell\) satisfies 
%     \begin{align*}
%       \limsup_{\ell} N_\ell n_\ell^{-\alpha} < \infty
%     \end{align*}
%     \item the estimates satisfy\footnote{Fourth moment integrability is needed to get a quantitative convergence rate for empirical second moments arising in bias correction, \S\ref{section:bc1} and onward.}
%     \begin{align*}
%         \expect \hat x_j - x_j &= O(h_\ell^{\beta})
%         \tag{bias}\\
%         \expect \left\|\hat x_j - \expect \hat x_j\right\|^4 &= O(h_\ell^{4\gamma})
%         \tag{fluctuation}
%         \\
%     \end{align*}
%   \end{enumerate}
% \end{defn}

We need some technical lemmas, proven in the Appendix, that estimate the asymptotics of locally dependent sums such as empirical gram matrix \(
\frac{1}{n'}
\sum_{j = 1}^{n'}
\hat \phi_j \hat \phi_j^\intercal
\) arising in least squares.
\begin{lem}
  \label{lem:consistent-sum}
  Let \(f\) be a smooth function with bounded first through third derivatives, and let \(\{\hat x_j\}_{j \in [1\ldots n']}\) come from a \((\beta, \gamma)\)-consistent filter.
  Then
  \begin{multline}
    \frac{1}{n'} \sum_{i = 1}^{n'} f(\hat x_i)
    - \frac{1}{n'} \sum_{i = 1}^{n'} f(x_i)
    \\
    = O_p(h^\beta + n^{-1/2} h^{- \frac{1}{2}\alpha + \gamma} + h^{2\gamma}).
  \end{multline}
  % In particular, if \(n h^{\alpha + 2\gamma} = O(1)\), then
  % \begin{multline}
  %   \frac{1}{n'} \sum_{i = 1}^{n'} f(\hat x_i)
  %   - \frac{1}{n'} \sum_{i = 1}^{n'} f(x_i)
  %   \\
  %   = O_p(h^\beta + n^{-1/2} h^{- \frac{1}{2}\alpha + \gamma}).
  %   \label{eq:consistent-sum-2}
  % \end{multline}
\end{lem}

\begin{fact}
  \label{fact:dependent-sum}
  For future use, let us distill the following intermediate result from the proof of Lemma~\ref{lem:consistent-sum}.
  When the terms are zero-mean and independent when separated by \(N \sim h^{-\alpha}\),
  \begin{align*}
    \frac{1}{n'} \sum_{i = 1}^{n'} O_p(1)
     & = O_p(n^{-1/2} h^{-\frac{1}{2}\alpha}).
  \end{align*}
\end{fact}

% 
% \begin{tabular}{llr}
%     {\bfseries Purpose} & \textbf{Gather} & to discover and experience\\
%     & \textbf{Grow} & to cultivate mutually \\
%     & \textbf{Go} & to participate in God's
% \end{tabular}
% 

% \begin{rem}
  Let us examine the three terms of Lemma 6, \(O_p(h^\beta + n^{-1/2} h^{- \frac{1}{2}\alpha + \gamma} + h^{2\gamma})\).
  The first, \(h^\beta\), is the Taylor expansion error of \(y\) and would matter if  \(y\) were measured without noise.
  The second, \(n^{-1/2} h^{- \frac{1}{2}\alpha + \gamma}\), refers to how the fluctuation induced by noise in \(y\) cancels out over large-sample averaging.
  The third, \(h^{2\gamma}\), which has no cancellation in \(n\), quantifies how \(\hat x_j \overset{\probability}{\rightarrow} \bar x_j\) implies \(\phi(\hat x_j) \overset{\probability}{\rightarrow} \phi(\bar x_j)\) by the Continuous Mapping Theorem.
  % The secondary conclusion \eqref{eq:consistent-sum-2} describes the large-sample regime
  % \(n h^{\alpha + 2\gamma} = O(1)\),
  % which we consider more interesting.
  % \todo{The following Lemma expands to a higher order an}
  The additive term \(h^{\beta}\) and the multiplicative factor \(h^{-\frac{1}{2} {\alpha}}\) are the price we pay for the convenience of plug-in estimation.
  All other things being equal, the Prediction Error Method is expected to achieve an error \(O_p(n^{-1/2})\) by classical large-sample theory.
% \end{rem}

\begin{lem}
  \label{lem:consistent-sum-2nd}
  Let \(f\) be a smooth function with bounded derivatives, and let \(\{\hat x_j\}_{j \in [1\ldots n']}\) come from a \((\beta, \gamma)\)-consistent filter.
  Then
  \begin{multline}
    \frac{1}{n'} \sum_{i = 1}^{n'}
    \sbr{
      f(\hat x_i)
      - \frac{1}{2}
      \partial_{\mu_1}\partial_{\mu_2} f(\hat x_i) \Sigma^{\mu_1\mu_2}
    }
    - \frac{1}{n'} \sum_{i = 1}^{n'} f(x_i)
    \\
    = O_p(h^\beta + n^{-1/2} h^{- \frac{1}{2}\alpha + \gamma} + h^{3\gamma}),
  \end{multline}
  where \(\Sigma^{\mu_1\mu_2} = \expect \Delta \tilde x_i^{\mu_1} \Delta \tilde x_i^{\mu_2}\).
\end{lem}

\begin{rem}
  The final \(h\) term in Lemma~\ref{lem:consistent-sum-2nd} is \(O_p(h^{3\gamma})\), which decays strictly faster than the corresponding term \(O_p(h^{2\gamma})\) in Lemma~\ref{lem:consistent-sum}.
  This observation is used for bias correction in Section~\ref{section:bc1}.
\end{rem}

In linear models, some kind of persistency of excitation (PE) is needed to quantify the asymptotic definiteness of the covariate gram matrix \cite{ljung_system_1999}.
Because of our dual limit \((n, h) \to (\infty, 0)\), our PE condition (Assumption~\ref{assum:persistency}) is stated in terms of a triangular array.
Whereas in DT LTI identification, the minimum singular value appearing in the above definition is a proxy for the asymptotic variance, we require it also to assume the task of regularizing the inversion of a noisy gram matrix:

\begin{lem}[Lipschitz continuity of matrix inversion]
  \label{lemma:matrix-inversion-continuity}
  Let \(\{\phi_{i, n'}\}_{i \in [1\ldots n'] }\) be a persistently exciting triangular array, and let \(\nu \rightarrow 0^+\).
  Then
  \begin{multline*}
    \del{
      \frac{1}{n'} \sum_{i = 1}^{n'} \phi_{i} \phi_{i}^\intercal
      + O_{\probability}(\nu)
    }^{-1}
    \\
    =
    \del{
      \frac{1}{n'} \sum_{i = 1}^{n'} \phi_{i} \phi_{i}^\intercal
    }^{-1}
    + O_{\probability}(\nu)
  \end{multline*}
\end{lem}

\section{Ordinary least squares (OLS)}
\label{section:ols}
Define for \(j \in [1 \ldots n']\):
\begin{subequations}
  \label{eq:ols-variables}
  \begin{align}
    \hat y_j    & = \hat x^{(m)}_j          \\
    \hat \phi_j & = \phi(\hat x_j^{(0\ldots m-1)}, t_j)
  \end{align}
\end{subequations}
and define the estimator:
\begin{align}
  \hat \theta_{\text{LS}}
   & = \del{
    \frac{1}{n'}
    \sum_{j = 1}^{n'}
    \hat \phi_j \hat \phi_j^\intercal
  }^{-1}
  \del{
    \frac{1}{n'}
    \sum_{j = 1}^{n'}
    \hat \phi_j \hat y_j
  }.
  \label{eq:least-squares-definition}
\end{align}
\begin{thm}[LS Consistency]
  \label{thm:ols-consistency}
  The LS estimator satisfies
  \begin{align*}
    \hat\theta_\text{LS} & = \theta_0 + O_p(h^\beta + n^{-1/2} h^{- \frac{1}{2}\alpha + \gamma} + h^{2\gamma}).
  \end{align*}
  % If \(\nu_\ell \to 0\), then \(\hat\theta_\text{LS}\) is a consistent estimator.
\end{thm}
\begin{pf}
  By Lemma~\ref{lem:consistent-sum},
  \begin{align}
    \begin{split}
      \hat \theta_{\text{LS}} & = \del{
        \frac{1}{n'}
        \sum_{j = 1}^{n'}
        \phi_j \phi_j^\intercal
        + O_p(h^\beta + n^{-1/2} h^{- \frac{1}{2}\alpha + \gamma} + h^{2\gamma})
      }^{-1}
      \\
                              & \quad \cdot
      \del{
        \frac{1}{n'}
        \sum_{j = 1}^{n'}
        \phi_j^\intercal y_j
        + O_p(h^\beta + n^{-1/2} h^{- \frac{1}{2}\alpha + \gamma} + h^{2\gamma})
      }
    \end{split}
    \\
    \intertext{By Lemma~\ref{lemma:matrix-inversion-continuity},}
    \begin{split}
      \hat \theta_{\text{LS}}
       & = \del{
        \frac{1}{n'}
        \sum_{j = 1}^{n'}
        \phi_j \phi_j^\intercal
      }^{-1}
      \del{
        \frac{1}{n'}
        \sum_{j = 1}^{n'}
        \phi_j y_j
      }
      \\
       & \quad + O_p(h^\beta + n^{-1/2} h^{- \frac{1}{2}\alpha + \gamma} + h^{2\gamma})
    \end{split}
    \\
     & = \theta_0
    + O_p(h^\beta + n^{-1/2} h^{- \frac{1}{2}\alpha + \gamma} + h^{2\gamma}).
  \end{align}
\end{pf}
% \begin{before}
%     however \ldots show results with terrible bias
% \end{before}

\section{Bias correction}
\label{section:bc1}
Using the measurement noise variance \(\sigma^2\) and the differentiation coefficient matrix \(D\) from Def~\ref{def:filtering}, define the following bias corrections to the quadratic sums that appear in \eqref{eq:least-squares-definition}.
% The bias corrections
%  for \(\frac{1}{n'}\sum_{j = 1}^{n'} \hat\phi_j \hat \phi_j^\intercal\) is
% are given by
% \clearpage
\begin{subequations}
  \label{eq:bcls-bias-estimate}
  \begin{align}
    \hat \Sigma_{\phi\phi}
     & = \frac{1}{n'} \sum_{j = 1}^{n'} \mathcal B \sbr{
      \hat\phi_j \hat \phi_j^\intercal
    } 
    \\
    \hat \Sigma_{\phi y}
     & = \frac{1}{n'} \sum_{j = 1}^{n'} \mathcal B\sbr{
      \hat\phi_j \hat y_j
    }.
  \end{align}
\end{subequations}
where \(\mathcal{B}\) is the operator
\begin{align}
  \mathcal B = \frac{1}{2} D_k^{d} D_k^{d'} \Sigma_\epsilon^{\ell, \ell'} \partial_{x^{\ell, (d)}} \partial_{x^{\ell', (d')}}.
  \label{eq:bcls-bias-operator}
\end{align}
The bias-corrected least squares estimator is given by
\begin{align}
  \begin{split}
    \hat\theta_{BC}
     & = \del{
      \frac{1}{n'}
      \sum_{j = 1}^{n'}
      \hat \phi_j \hat \phi_j^\intercal
      -
      \hat \Sigma_{\phi\phi}
    }^{-1}
    \\
     & \quad\cdot
    \del{
      \frac{1}{n'}
      \sum_{j = 1}^{n'}
      \hat \phi_j \hat y_j
      -
      \hat \Sigma_{\phi y}
    }.
  \end{split}
\end{align}

\begin{thm}[BC Consistency]
  \label{thm:bcls-consistency}
  The BC estimator satisfies
  \begin{align*}
    \hat\theta_\text{BC} & = \theta_0 + O_p(h^\beta + n^{-1/2} h^{- \frac{1}{2}\alpha + \gamma} + h^{3\gamma}).
  \end{align*}
\end{thm}
\begin{pf}
  See the proof of Theorem~\ref{thm:ols-consistency}, but instead of Lemma~\ref{lem:consistent-sum}, use Lemma~\ref{lem:consistent-sum-2nd} on the bias-corrected sums.
\end{pf}

\begin{rem}
  Compare this result to the LS consistency result (Thm.~\ref{thm:ols-consistency}).
  The common terms are differentiation bias \(h^\beta\), which would be a bias term even if \(y\) were discretely without noise; and first-order fluctuation \(n^{-1/2} h^{- \frac{1}{2}\alpha + \gamma}\), which is the linearized effect of observation noise.
  The third term is the nonlinear effect of observation noise.
  In the BC estimator, this term is an order of magnitude smaller in \(h\).
\end{rem}

% The BCLS estimator has the same performance to leading order, but its limit \((n, h) \to (\infty, 0)\) is more robust to \(h\).
% It requires \(nh^{\alpha + 4\gamma}=O(1)\), which allows for asymptotically smaller \(n\) or larger \(h\) than the OLS hypothesis \(nh^{\alpha + 2\gamma} =O(1)\).

\begin{rem}
  % The effect of bias correction is that 
  The matrices \(\hat\Sigma_{\phi\phi}, \hat\Sigma_{\phi y}\) can be viewed as a perturbative nonlinear generalization of ``bias compensation'' in least squares linear system identification \cite[Chapter 7]{soderstrom_errors--variables_2018} \cite{hong_accuracy_2007,mejari_bias-correction_2018,mejari_recursive_2020,piga_bias-corrected_2014}.
  For a class of nonlinearitiess including polynomial \cite{piga_bias-corrected_2014}, knowledge of the noise distribution allows for exact bias correction by deconvolution \cite[\S3]{schennach_recent_2016} \cite{fazekas_asymptotic_1997,baran_consistent_2000}.
  % On the other hand, \(\hat \Sigma_{\phi y}\), which corrects for correlation (endogeneity) between the regressors and predictors, may be the first correction of its kind.
\end{rem}

\section{An instrumental variables method}
\label{section:iv}
Recall again that the least squares estimator can be written,
\begin{align}
  \hat \theta_\text{LS}
   & =\del{
    \frac{1}{n'}
    \sum_{j = 1}^{n'}
    \hat \phi_j \hat \phi_j^\intercal
  }^{-1}
  \del{
    \frac{1}{n'}
    \sum_{j = 1}^{n'}
    \hat \phi_j \hat y_j
  }.
  \intertext{In a traditional instrumental variables setup, one replaces some of the occurences of \(\hat \phi_i\) with another series of vectors \(\psi_i\), known as instruments, resulting in an expression along the lines of}
   & \phantom{{}={}} \del{
    \frac{1}{n'}
    \sum_{j = 1}^{n'}
    \psi_j \hat \phi_j^\intercal
  }^{-1}
  \del{
    \frac{1}{n'}
    \sum_{j = 1}^{n'}
    \psi_j \hat y_j
  }.
\end{align}
Instrumental variables estimation originated in the social sciences to deal with estimation of the linear regression \(\expect (y \mid x) = \beta^\intercal x\) when the explanatory variable \(x\) follows a random design correlate with noise in \(y\) \cite{angrist_mostly_2009,bollen_instrumental_2012,pischke_lecture_nodate,schennach_recent_2016,amemiya_advanced_1985}.
% In spite of their higher variance, IV estimators are often preferred due to their robustness to bias.

An IV estimator has low bias if \(\psi_j\) and \(\hat \phi_j\) are uncorrelated as random variables, and low variance if they are strongly correlated across \(j\).
% The former criterion eliminates bias arising from the quadratic nonlinearity in the normal equations:
% \begin{align}
%   \frac{1}{n'}
%   \sum_{j = 1}^{n'}
%   \psi_j \hat \phi_j^\intercal
%    & \to \frac{1}{n'}
%   \sum_{j = 1}^{n'}
%   \expect  \psi_j  \expect \hat \phi_j^\intercal
%   \intertext{and}
%   \frac{1}{n'}
%   \sum_{j = 1}^{n'}
%   \psi_j^\intercal \hat y_j
%    & \to
%   \frac{1}{n'}
%   \sum_{j = 1}^{n'}
%   \expect \psi_j^\intercal \expect \hat y_j.
% \end{align}
% The latter criterion is an IV version of persistency of excitation.
% % , and reduces the first-order noise sensitivity of \(\hat \theta_\text{IV}\).
Our IV-inspired estimator achieves both criteria by \(\psi_j = \tilde \phi_j\), a second estimate of \(\phi_j\) based on disjoint data points and thus stochastically independent but similarly distributed to \(\hat \phi_j\).
This can be interpreted as a higher-order, smoothed version of lagged instrumental variables \cite{bollen_instrumental_2012}.
% In this section, we show how the additive white noise assumption allows us to construct instruments \(\tilde \phi_j\) that are ideal in the following sense:
% while

Let \(\tilde D\) and \(D\) be the coefficient matrices for two \((\beta, \gamma)\)-consistent overlapping filter schemes having identical \(N\) and \(t_j\).
% Having the ability to customize the output time \(t_j\) necessitates the ``accurate'' filter presented in Section~\ref{section:filtering}.\todo{impacted by filter}
Furthermore, we require that the even columns of \(\tilde D\) be zero and that the odd columns of \(D\) be zero: this ensures that each random data point \(z_i\) is used either in the \(\tilde D\) output or the \(D\) output, but not both.

For \(j \in [1 \ldots n']\), let the two filters' outputs be written as \(\hat x\) and \(\tilde x\), respectively, according to Def.~\ref{def:filtering}:
\begin{align}
  \hat x^{\ell, (d)}_j
   & = \sum_{k = 1}^N D^d_k z_{n_j + k}^\ell
  \\
  \tilde x^{\ell, (d)}_j
   & = \sum_{k = 1}^N \tilde D^d_k z_{n_j + k}^\ell
\end{align}
In the spirit of \eqref{eq:ols-variables}, define bias-corrected\footnote{Bias correction at this stage is a technical requirement, but in practice these biases, which scale as the second derivatives of \(\phi\), are often significantly smaller than \(\phi\) and can be omitted if e.g.~\(\sigma^2\) is not known. If we are identifying an LTI system, then they are in fact zero.} \(\hat \phi_j\) and \(\tilde \phi_j\), independent approximations to \(\phi_j\); and \(\hat y_j\) and \(\tilde y_j\), independent approximations to \(y_j\):
\begin{subequations}
  \label{eq:iv-variables}
  \begin{align}
    \hat y_j      & = \hat x^{(m)}_j
    \\
    \tilde y_j    & = \tilde x^{(m)}_j
    \\
    \hat \phi_j   & = (1 - \mathcal B_D)\phi(\hat x_j^{(0\ldots m-1)}, t_j) 
    \\
    \tilde \phi_j & = (1 - \mathcal B_{\tilde D}) \phi(\tilde x_j^{(0\ldots m-1)}, t_j)
  \end{align}
\end{subequations}
where \(\mathcal B_D\) and \(\mathcal B_{\tilde D}\) specify which \(D\) is used to define the generic bias-correction operator \eqref{eq:bcls-bias-operator}.
The IV-inspired estimator is defined as
\begin{multline}
  \label{eq:iv-estimator}
  \hat \theta_\text{IV}
  = \sbr{
    \frac{1}{2n'} \sum_{j = 1}^{n'}
    \del{
      \hat \phi_j \tilde \phi_j^\intercal
      +
      \tilde \phi_j \hat \phi_j^\intercal
    }
  }^{-1}
  \\
  \cdot \sbr{
    \frac{1}{2n'} \sum_{j = 1}^{n'}
    \del{
      \tilde \phi_j \hat y_j
      +
      \hat \phi_j \tilde y_j
    }
  }.
\end{multline}
\begin{thm}[IV Consistency]
  \label{thm:iv-consistency}
  The IV estimator satisfies
  \begin{align*}
    \hat\theta_\text{IV} & = \theta_0 + O_p(h^\beta + n^{-1/2} h^{- \frac{1}{2}\alpha + \gamma} + h^{3\gamma}).
  \end{align*}
\end{thm}
% \section{Proof of Thm~\ref{thm:iv-consistency}}
\begin{pf}
  % The proof uses similar techniques to the proof of Lemma~\ref{lem:consistent-sum-2nd}.
  Let \(\eta_j = \hat x_j - \expect \hat x_j\) and \(\varepsilon_j = \tilde x_j - \expect \tilde x_j\).
  Let us expand \(\hat \phi_j\) and \(\tilde \phi_j\) as
  \begin{align}
    \begin{split}
      \hat \phi_j
       & = \phi(\expect \hat x_j)
      + \partial_{\mu_1} \phi(\expect \hat x_j) \eta_j^{\mu_1}
      \\
       & \quad
      + \frac{1}{2} \partial_{\mu_1} \partial_{\mu_2} \phi(\expect \hat x_j)
      \sbr{\eta_j^{\mu_1} \eta_j^{\mu_2}
        - \expect \eta_j^{\mu_1} \eta_j^{\mu_2}}
      \\
       & \quad + O_{p}(h^{3\gamma})
    \end{split}
    \\
    \begin{split}
      \tilde \phi_j
       & = \phi(\expect \tilde x_j)
      + \partial_{\mu_1} \phi(\expect \tilde x_j) \varepsilon_j^{\mu_1}
      \\
       & \quad
      + \frac{1}{2} \partial_{\mu_1} \partial_{\mu_2} \phi(\expect \tilde x_j)
      \sbr{\varepsilon_j^{\mu_1} \varepsilon_j^{\mu_2}
        - \expect \varepsilon_j^{\mu_1} \varepsilon_j^{\mu_2}}
      \\
       & \quad + O_{p}(h^{3\gamma})
    \end{split}
  \end{align}
  % todo rewrite this section using A, B formalism
  We next analyze the outer product \(\hat \phi_j \tilde \phi_j^\intercal\) using indices \(\nu, \nu'\).
  Rather than enumerate all sixteen terms in this estimate, let us summarize the leading order terms in order of polynomial degree in \(\eta_j\) and \(\varepsilon_j\).
  \begin{enumerate}
    \setcounter{enumi}{-1}
    \item  By Lipschitz continuity, \(\phi_j^\nu(\expect \hat x_j) \phi_j^{\nu'}(\expect \tilde x_j)
          =
          \phi_j^\nu(x_j) \phi_j^{\nu'}(x) + O(h^\beta)\)
    \item
          \(\phi_j^\nu(\expect \hat x_j) \partial_{\mu_1} \phi_j^{\nu'}(\expect \tilde x_j) \varepsilon_j^{\mu_1}\) and its counterpart
          are \(O_p(h^{\gamma})\) with zero mean.
          %  and therefore exhibits cancellation when summed over \(j \in [n']\).
    \item
          \begin{enumerate}
            \item \(\phi_j^\nu(\expect \hat x_j)\frac{1}{2} \partial_{\mu_1} \partial_{\mu_2} \phi(\expect \tilde x_j)
                  \sbr{\varepsilon_j^{\mu_1} \varepsilon_j^{\mu_2}
                    - \expect \varepsilon_j^{\mu_1} \varepsilon_j^{\mu_2}}\) and its counterpart
                  are \(O_p(h^{2\gamma})\) with zero mean.
            \item
                  \label{eq:1-by-1-cross-term}
                  \(\partial_{\mu_1} \phi^{\nu}(\expect \hat x_j) \eta_j^{\mu_1}
                  \partial_{\mu_2} \phi^{\nu'}(\expect \tilde x_j) \varepsilon_j^{\mu_2}
                  \) and its counterpart
                  are \(O_p(h^{2\gamma})\) with zero mean.
            \item All of the remaining terms are \(O_p(h^{3\gamma})\).
          \end{enumerate}
  \end{enumerate}

  When this investigation is also applied to \(\tilde \phi_j \hat \phi_j^\intercal\), \(\tilde \phi_j\hat y_j\), and \(\hat \phi_j \hat \tilde y_j\), we conclude:
  \begin{subequations}
    \begin{align}
      \hat \phi_j \tilde \phi_j^\intercal
      +
      \tilde \phi_j \hat \phi_j^\intercal
       & = 2 \phi(x_j) \phi(x_j)^\intercal + O_p(h^\beta + h^\gamma + h^{3\gamma})
      \\
      \tilde \phi_j \hat y_j
      +
      \hat \phi_j \tilde y_j
       & = 2 \phi(x_j) y_j + O_p(h^\beta + h^\gamma + h^{3\gamma})
    \end{align}
  \end{subequations}
  where the \(h^\gamma\) term exhibits cancellation according to Fact~\ref{fact:dependent-sum}.
\end{pf}
\begin{rem}[IV compared to OLS and BCLS]
  The cross term \ref{eq:1-by-1-cross-term} in the above proof
  has zero mean because \(\eta_j\) and \(\varepsilon_j\) have zero mean and are independent, thus uncorrelated.
  In the OLS estimator, \(\varepsilon_j\) would be a.s.~equal to \(\eta_j\), which leads to an upward bias in gram matrix of \(\hat \Phi\) and therefore a downward bias in \(\hat\theta_{\text{LS}}\).
  Whereas the BCLS method estimates it, the IV estimator allows it to cancel itself over \(j \in [n']\).
\end{rem}
% TODO 11/9/24 finish proof here
% \color{black!10!white}
\section{Filtering}
\label{section:filtering}
We present one construction that meets the stipulations of \((\beta, \gamma)\)-consistent filtering (Def.~\ref{def:filtering}).

\begin{rem}
  Our analysis on filtering concerns asymptotic rates \(N\sim h^{-\alpha}\), constants ignored, as \(h\to 0\).
  In practical applications the window size \(N\) will need to be selected subjectively or based on a criterion such as cross-validation---a heavily investigated question in applications for estimation and inference at a single point \cite{imbens_optimal_nodate,fan_variable_1992,fan_data-driven_1995,rupperts_effective_1995,jones_brief_1996,gijbels_local_1998,katkovnik_adaptive_1998,prewitt_bandwidth_2006,raykar_fast_2006,menon_sure-optimal_2013}.
\end{rem}

The accurate filter is constructed in Appendix~\ref{section:accurate-time}, where Lemmas~\ref{lem:acc-diff-bias} and \ref{lemma:acc-diff-variance} yield the following:
\begin{thm}
  \label{thm:accurate-consistency}
  The \(p\)-accurate filter is \((\beta, \gamma)\) consistent if
  \(\frac{2m}{2m + 1} < \alpha < 1\),
  %  and \(h_\ell^{\delta} n_\ell^{-\alpha} = O(1)\),
  with constants
  \begin{align*}
    % \alpha &= \\
    \beta  & = (p - m)(1 - \alpha),           \\
    \gamma & = \frac{(2m + 1)\alpha - 2m}{2}.
  \end{align*}
\end{thm}

% \begin{exmp}[Asymptotically optimal \(\delta\)]
%     Recall from Lemma~\ref{lem:consistent-sum} that the error rate is \(\nu = h^\beta + n^{\frac{\alpha - 1}{2}} h^\gamma\).
%     In the accurate filter, we have \(n^\alpha \sim h^{\delta}\), so
%     \begin{align}
%         \nu &= h^{(p -m)(1 + \delta)} + h^{\frac{\delta - 1}{2} + \frac{2m (1 + \delta) + \delta}{2}}.
%         \intertext{Balancing, we get the optimal rate}
%         % 
%     \end{align}
% \end{exmp}

% \subsection{Optimal filtering}

\section{Numerical example: van der Pol oscillator}
\label{sec:example-vdp}
We demonstrate our estimators on the following van der Pol oscillator, observed with additive white Gaussian noise:
\begin{align}
  \ddot y(t) & = \theta_1 (1 - y^2) \dot y(t) + \theta_2 y(t),
  \quad 0 \leq t \leq T = nh.
\end{align}
Pertinent constants relating to the simulated experiment and estimators are available in Table~\ref{table:constants}.
In order to assess the sampling distributions of these estimators, we simulated 10,000 realizations of the noisy trajectory and applied each estimator to each realization.

\begin{figure}
  \includegraphics[width=\linewidth]{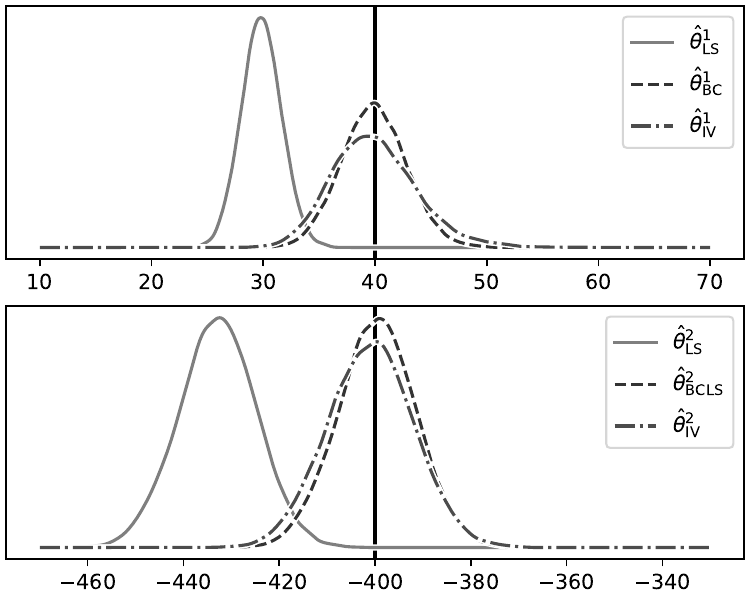}
  \caption{
    \label{fig:theta_results}
    Kernel density estimate of the sampling distribution of the estimated \(\theta = (\theta^1, \theta^2)\) of \S\ref{sec:example-vdp} under different pairings of differentiation and regression methods. True values indicated by a vertical line.
  }
\end{figure}

We present the sampling distributions in Figures~\ref{fig:theta_results}.
We present the summary statistics, normalized by the true parameter magnitudes in Tables~\ref{table:theta_1_risk} and \ref{table:theta_2_risk}.

From visual inspection of the sampling distributions and from consulting the summary statistics:
\begin{enumerate}
  \item The plain LS estimator has the greatest bias regardless of the underlying differentiation filter. We attribute this to the nonlinear effect of observation noise, which is the \(O_p(h^{2\gamma})\) term in the OLS consistency result (Thm.~\ref{thm:ols-consistency}).
  \item We attribute the smaller bias of the BCLS and IV estimators to the reduction of the \(O_p(h^{2\gamma})\) term to \(O_p(h^{3\gamma})\).
        % \item The variability of these estimators comes from the \(n^{-1/2} h^{- \frac{1}{2}\alpha + \gamma}\) terms in the consistency results.
  % \item The bias in the orthogonal differentiation filter does not go away with the nonlinear correction for observation noise.
  %       We attribute it to the \(O_p(h^\beta)\) term for oversmoothing in the local polynomial step.
  % \item The orthogonal differentiation filter results in estimates having the least error on average. (We found in other numerical studies that this phenomenon is not highly sensitive to \(N\) or \(p\).)
  %       We attribute this chiefly to the off-diagonal bias cancellation property discussed in Remark~\ref{rem:filters-compared}.
  %       This very particular utility of orthogonal filters has seemingly not been harnessed in the literature on signal processing for system identification.
  %       This comparison also demonstrates that when the data is noisy, some smoothing can be desirable even when it comes  at the expense of Taylor approximation accuracy.
        % \item In keeping with econometric wisdom, the IV estimators have higher variance than the LS-based estimators.
\end{enumerate}

% \begin{figure}
%   \includegraphics[width=\linewidth]{graphics/θ_2_estimates}
%   \caption{
%     \label{fig:theta_2_results}
%     Kernel density estimate of the sampling distribution of the estimated \(\theta_2\) under different pairings of differentiation and regression methods. True value indicated by a vertical line.
%   }
% \end{figure}

\begin{table}
  \begin{center}
    \begin{tabular}{llrrr}
      Regression & bias (\%) & std (\%) & RMSE (\%)
      \\\hline
      LS & -25.39 & 4.49 & 25.78\\
      BCLS & -0.21 & 7.26 & 7.26\\
      IV & -0.07 & 9.61 & 9.61\\
      \hline
    \end{tabular}
  \end{center}
  \caption{\label{table:theta_1_risk}%
  Statistics from the sampling distribution of estimators for \(\theta_1\) of \S\ref{sec:example-vdp}}
\end{table}

\begin{table}
  \begin{center}
    \begin{tabular}{llrrr}
      Regression & bias (\%) & std (\%) & RMSE (\%)
      \\\hline
      LS & -8.22 & 2.01 & 8.46\\
      BCLS & 0.10 & 2.02 & 2.02\\
      IV & -0.23 & 2.24 & 2.26\\
      \hline
    \end{tabular}
  \end{center}
  \caption{\label{table:theta_2_risk}%
  Statistics from the sampling distribution of estimators for \(\theta_2\) of \S\ref{sec:example-vdp}.}
\end{table}

\begin{table}
  \begin{center}
    \begin{tabular}{lrr}
      Variable                 & Meaning                      & Value          \\\hline
      \(n\)                    & number of measurements       & 2000           \\
      \(h\)                    & sampling period              & 1/2000         \\
      \(T\)                    & trajectory duration          & 1              \\
      \(m\)                    & highest derivative needed    & 2              \\
      \(p\)                    & filter order & 6              \\
      \(N\)                    & filter window length         & 50             \\
      \(\sigma^2\)             & noise variance               & 0.01           \\
      \((y(0), \dot y(0))\)    & initial condition            & \((0, 0.001)\)
      \\
      \((\theta_1, \theta_2)\) & true parameter               & \((40, -400)\)
      \\
      \hline
    \end{tabular}
  \end{center}

  \caption{%
    \label{table:constants}%
    Details of the example problem in \S\ref{sec:example-vdp}.
  }
\end{table}

\section{Numerical example: Lorenz system}
\label{sec:example-lorenz}
\begin{figure}
  \includegraphics[width=\linewidth]{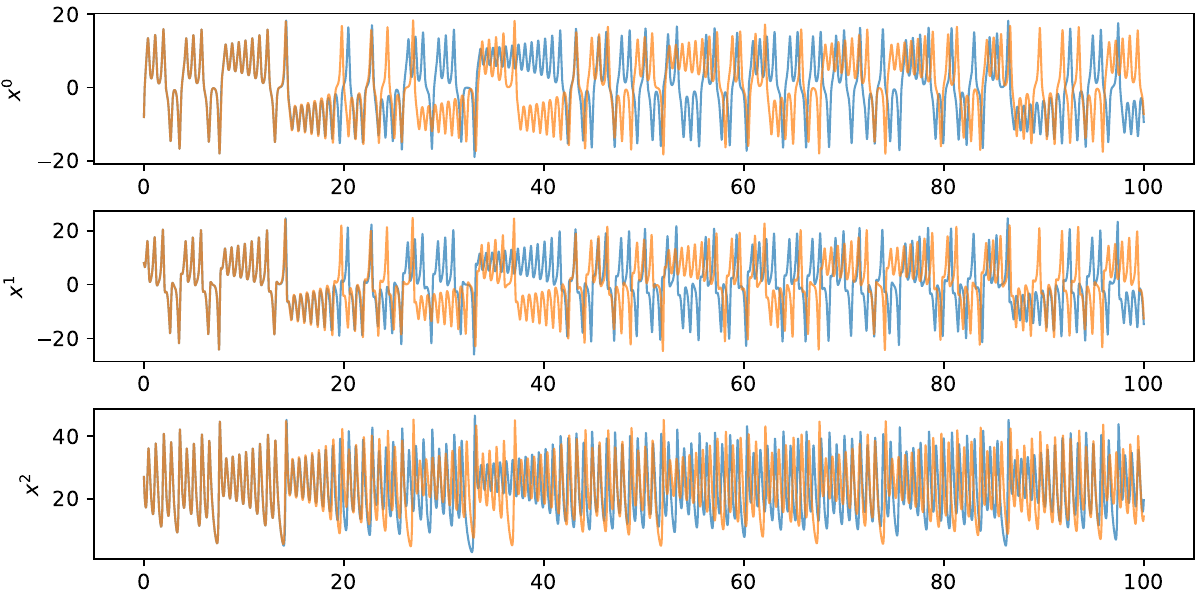}
  \caption{
    \label{fig:lorenz-pem}
    Two numerically indistinguishable solutions fo the Lorenz system initial value problem.
  }
\end{figure}

This section applies all three estimators to the three-dimensional Lorenz system for \(x :[0, T] \to \mathbb{R}^3\), specified as
\begin{subequations}
\label{eq:lorenz}
\begin{align}
  \dot x &= \phi(x)^\intercal A_0.
  \intertext{We replicate \cite[\S4.2]{brunton_discovering_2016} by}
  \phi(x) &= \begin{pmatrix}
    x^1\\ x^2\\ x^3\\ x^1 x^2\\ x^1 x^3
  \end{pmatrix}
  \\
  A_0 &= 
  \begin{pmatrix}
-10 & 28 & 0 \\
10 & -1 & 0 \\
0 & 0 & -8/3 \\
0 & 0 & 1 \\
0 & -1 & 0
\end{pmatrix}
\end{align}
\end{subequations}
and using the same initial conditions and measurement times.
Only \(x\) is measured, and with white Gaussian noise.
The parameter of interest is the matrix \(A_0\).
Details are listed in Table~\ref{table:lorenz-constants}.

First we establish that the Prediction Error Method with a state space model is ineligible for this task due to the chaotic dynamics.
Compare the two state-space trajectories in Fig~\ref{fig:lorenz-pem}.
Both are numerical solutions to the initial value problem \eqref{eq:lorenz} using the true initial condition and \(A_0\).
They are integrated using a 5th order implicit scheme with an adaptive step size.
One of them (which we use hereafter as training data) is initialized with a step size of 0.001, replicating \cite[\S4.2]{brunton_discovering_2016}, and the other is initialized with a step size of 0.002.
It does not matter which is which.
All state space predictions past \(t \approx 20\) are effectively pseudorandom.

\begin{table}
  \begin{center}
    \begin{tabular}{lrr}
      Variable                 & Meaning                      & Value          \\\hline
      \(n\)                    & number of measurements       & 100000           \\
      \(h\)                    & sampling period              & 0.001         \\
      \(T\)                    & trajectory duration          & 100              \\
      \(m\)                    & highest derivative needed    & 1              \\
      \(p\)                    & filter order & 50              \\
      \(N\)                    & filter window length         & 200             \\
      \(\sigma^2\)             & noise variance               & 0.1, 10           \\
      \((x^1(0), x^2(0), x^3(0))\)    & initial condition            & \((-8, 8, 27)\)
      \\\hline
    \end{tabular}
  \end{center}

  \caption{%
    \label{table:lorenz-constants}%
    Details of the example problem in \S\ref{sec:example-lorenz}.
  }
\end{table}

\begin{figure*}
  \includegraphics[width=\linewidth]{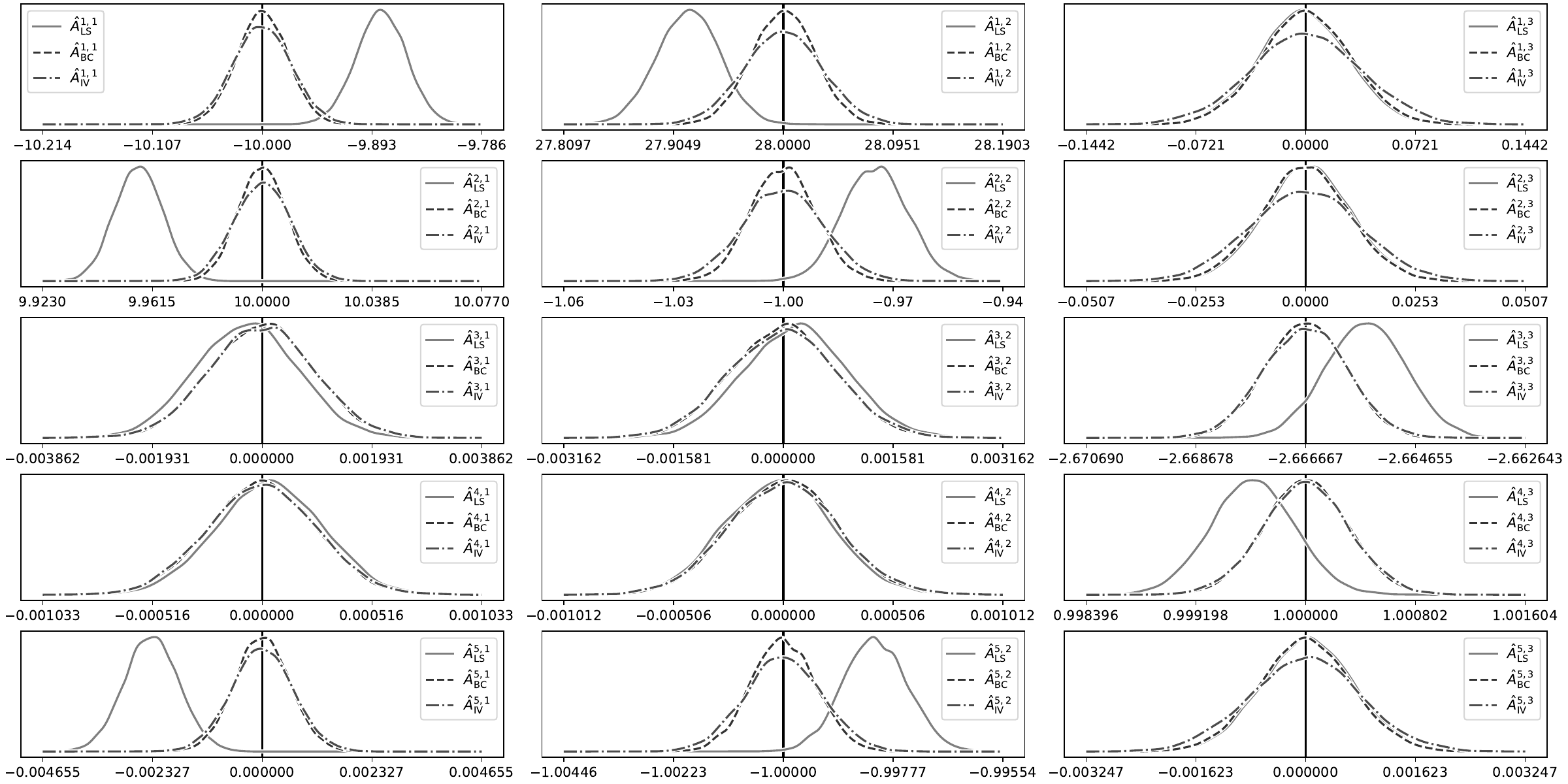}
  \caption{
    \label{fig:lorenz-small-variance}
    Kernel density estimate of the sampling distribution of the estimated \(A_0\) of \S\ref{sec:example-lorenz} with \(\sigma^2=0.1\).
    True values indicated by vertical lines.
  }
\end{figure*}
\begin{figure*}
  \includegraphics[width=\linewidth]{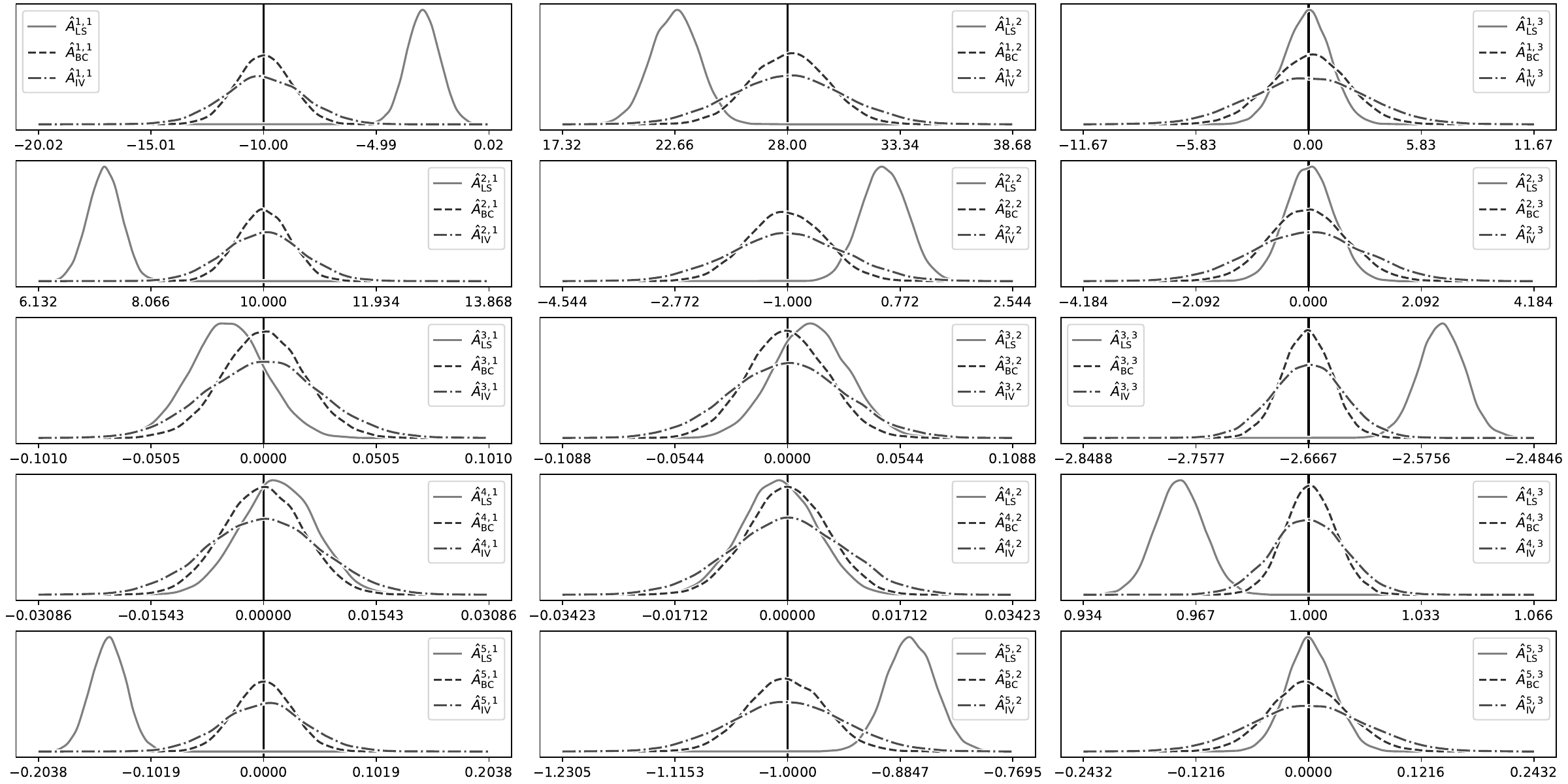}
  \caption{
    \label{fig:lorenz-big-variance}
    Kernel density estimate of the sampling distribution of the estimated \(A_0\) of \S\ref{sec:example-lorenz} with \(\sigma^2=10\).
    True values indicated by vertical lines.
  }
\end{figure*}

\begin{table}
  \begin{center}
    \begin{tabular}{llrrr}
      \(\sigma^2\) & method & bias (\%) & RMSD (\%) & RMSE (\%) 
      \\
      \hline
     \(0.1\) & LS & 0.47811 & 0.15584 & 0.50287\\
    \(0.1\) & BC & 0.00098 & 0.15690 & 0.15690\\
    \(0.1\) & IV & 0.00126 & 0.18972 & 0.18972\\
      \hline
      \(10\) & LS & 29.83869 & 6.08035 & 30.45190\\
    \(10\) & BC & 0.19944 & 9.83263 & 9.83465\\
    \(10\) & IV & 0.01395 & 14.56300 & 14.56300\\
      \hline
    \end{tabular}
    \caption{%
      \label{table:lorenz-risks}%
      Bias, RMS deviation from the mean (in operator norm), and RMSE (in operator norm) of the estimated \(A_0\) of \S\ref{sec:example-lorenz} with \(\sigma^2=0.1\) and \(\sigma^2=10\).
      % All scaled by operator norm of \(A_0\).
      }
  \end{center}
\end{table}
For \(\sigma^2 \in \{0.1, 10\}\), we simulated 10,000 realizations of the noisy trajectory and applied each estimator to each realization.
We show the marginal distributions of the elements of \(\hat A\) in Figure~\ref{fig:lorenz-small-variance} (\(\sigma^2=0.1\)) and Figure~\ref{fig:lorenz-big-variance} (\(\sigma^2=10\)).
The LS estimator is biased toward zero, as our theoretical narrative predicts.
In conjunction with the statistics in Table~\ref{table:lorenz-risks}, we can see that at both noise scales, the BC and IV estimators achieve a nearly 50x reduction in bias at the cost of a 1.5x increase in fluctiation, netting an overall threefold reduction in RMSE.
While the BC estimator appears to have a lower RMSE risk, the IV estimator is more robust to the prior information about \(\sigma^2\).
(In this example, the function \(\phi\) happens to be harmonic, which means that the IV estimator's embedded bias-correction is zero.)

\section{Conclusion}
% Despite its popularity in applied engineering, nonlinear continuous-time system identification lacks direct methods.
We show that a least squares method can be asymptotically consistent but biased for realistic sampling conditions, and that this bias can be usefully eliminated to second order by two bias correction methods.
% Pre-smoothing before estimating a continuous-time autoregression involves a bias-variance tradeoff in the choice of smoothing scale \(N\) and polynomial order \(p\).
Our examples show that across a range of problems and signal-to-noise ratios, the BC and IV estimators dominate LS in terms of bias and quadratic risk.

\begin{ack}                               % Place acknowledgements
This work was supported by the National Science Foundation
CAREER Program (Grant No.~2046292).  % here.
\end{ack}

\appendix
\section{Related work}
\label{section:related}
\subsection{Bias}
Measurement error leads to a bias when the estimator is nonlinear in the measurements.
For example, if the estimator is \(f(z) = z^2\), then \(\expect f(z) - f(\expect z) = \sigma^2\) for \(z \sim \mathcal{N}(0, \sigma^2)\).
Abstract principles of our bias correction can be found in the statistics and econometrics literature \cite[Chapter 10]{wasserman_all_2006}, \cite{schennach_recent_2016,li_update_2024}.

\cite{piga_bias-corrected_2014} and \cite{mejari_bias-correction_2018} work out bias-corrected least squares (\S\ref{section:bc1}) for discrete-time systems.
Under certain assumptions on the measurement noise and nonlinearity, it can be possible to correct noise-induced bias exactly.

An analogous bias mechanism appears in the related problem of Errors-in-Variables system identification, in which the both the input and output of a linear system are measured with error.
Old and new methods for estimation are reviewed in \cite{soderstrom_errors--variables_2018}.
In particular, algebraic bias compensation methods for linear systems are analyzed in \cite{hong_accuracy_2007} and \cite{hong_relations_2009}.

\cite{fan_estimation_1999} describes a method for reducing bias in continuous-time autoregressive systems by using a pair of uncorrelated filters similar to our IV filter design (\S\ref{section:iv}).
Another approach to orthogonal filtering from the frequency-domain algebraic perspective is multiple prefiltering \cite{mahata_identification_2002}.

The term bias correction sometimes refers to concerns other than noise-induced bias.
Works such as \cite{pan_consistency_2020,gonzalez_consistency_2024} examine the systematic bias resulting from incompatible discretizations of a continuous-time LTI system.
This is related to bias arising from the Taylor approximation of a continuous-time response in \cite{soderstrom_least_1997}.

Bias can also arise when a direct method that is consistent for systems under random excitation is applied to a system under closed-loop control.
Viewing markets as a closed-loop control system is a motivation for  instrumental variables in structural econometrics.\footnote{For example, in a \emph{static} simultaneous equations model derived from a competitive equilibrium of supply and demand, supply shocks can be used as instruments to estimate supply elasticity \cite{colina_california_2024}. For an example of a closed-loop dynamic model, see \cite{gafarov_wild_2024}.}
A recent application of IV to a continuous-time model is \cite{pan_consistency_2022}.
A BCLS-style treatment may be found in \cite{mejari_bias-correction_2018}.

\subsection{Process vs.~output noise models}
\label{subsection:process-vs-output-noise}
In linear system identification and continuous time in particular, it makes a big difference whether we model our system uncertainty by measurement noise (deterministic evolution, noisy measurements) \cite{niethammer_parameter_2001,unbehauen_identification_1997,gonzalez_continuous-time_2022}
or process noise (random evolution, clean measurements) \cite{fan_estimation_1999,dinh-tuan_pham_estimation_2000,soderstrom_least_1997,soderstrom_bias-compensation_1997}.
By stochastic continuity, process noise vanishes as \(h\to 0\); for example, if \(B_t\) is a Brownian motion, then \(B_{t+h} - B_t = O_{\probability}(h^{1/2})\).
In process noise models, one encounters guarantees such as \(\lim_{h \to 0^+} \operatornamewithlimits{plim}_{n \to \infty} \text{error}(n, h) = 0\), where the proof technique is to appeal to ergodicity to compute the inner \(\operatornamewithlimits{plim}\) and then use Taylor expansion techniques to assess the outer limit.
These guarantees are strictly weaker than the consistency claims in our work [Theorems \ref{thm:ols-consistency}, \ref{thm:bcls-consistency}, and \ref{thm:iv-consistency}].

In the output error literature, on the other hand, noise-induced bias does not automatically disappear in the limit, and must be cancelled by careful estimator design.
Ignoring its effect leads to e.g.~the asymptotically inconsistent State Variable Filtering method for continuous-time linear systems.

\subsection{Differentiation of noisy data}
% book
Conventional wisdom in CT system identification holds that one must ``avoid the direct differentiation of noisy data''  \cite[p.~vi]{wang_system_2012}.
% Our paper defies this advice and is not the first to do so.
We are not the first to disregard this advice.
\cite{van_breugel_numerical_2020} is a practically-oriented review of algorithms for differentiating noisy data.
In the chemistry literature, the use of local polynomial fits for smoothing and differentiation is called Savitzky-Golay filtering \cite{barak_smoothing_2002}.

State Variable Filtering (SVF) \cite[Chapter 1]{garnier_identification_2008} preconditions the CT LTI model
\begin{align}
  A(p) \expect y          & = B(p) u
  \label{eq:svf-unconditioned}
  \intertext{to}
  D(p)\inv A(p) \expect y & = D(p)\inv B(p) u,
  \label{eq:svf-conditioned}
\end{align}
where \(\expect\) denotes expectation over square-integrable measurement noise, \(A\) and \(B\) are unknown degree-\(m\) polynomials in the differentiation operator \(p = \od{}{t}\), \(y\) is output, \(u\) is input, and \(D\) is a known stable polynomial of degree at least \(m\).
It follows from commuting linear operators or manipulating transfer functions in Laplace domain that \eqref{eq:svf-conditioned} is an equivalent model to \eqref{eq:svf-unconditioned}.
If \(y\) is observed over a long, persistently exciting trajectory \emph{without noise} then \eqref{eq:svf-conditioned} may be estimated consistently by least squares regression.
However, in the presence of measurement noise, the SVF is asymptotically biased.

An example of SVF for a first-order system is the approximation of Laplace transfer functions \(p \approx \hat p = \frac{p}{\tau p + 1}\), where \(\tau\) is a small positive constant.\footnote{An approximation also used in control, see \cite{marchi_dirty_2022,kuang_dirty_2023}}
When viewed as an meromorphic function in the symbol \(p\), \(\hat p\) is a first-order Pad\'e approximant of \(p\).
The order of Pad\'e approximation in the operator/Laplace domain corresponds to the order of accuracy in time domain.
If for some integer \(d>0\), coefficients \(D^d_k\) are chosen such that the Laurent series in the time delay operator \(z = e^{-hp}\)
\begin{align}
  p^d & \approx \sum_{k} D^d_k z^k
  \label{eq:approx-deriv-frequency-domain}
\end{align}
is a Pad\'e approximant of degree \(q\), it is equivalent to requiring that the coefficients \(D^d_k\) correctly differentiate time-domain polynomials of degree up to \(q\) (``natural conditions''  \cite{soderstrom_least_1997}).

The works \cite{soderstrom_least_1997,dinh-tuan_pham_estimation_2000} employ approximations where the support of \(D\) is reduced to as small as possible.
These differentiation rules are identical to finite difference stencils as used in numerical solutions of differential equations, which correspond to local polynomial \emph{interpolation} of \(y\) \cite{boyd_chebyshev_nodate}.
But because the local polynomial interpolation coefficients blow up as \(h^{-d}\) as \(h \to 0\),
we mitigate this blowup by expanding the support (bandwidth) of \(D\) as \(h\to 0\) so as to allow for local polynomial  \emph{regression} (LPR) \cite{fan_local_2003}.
LPR is used in econometrics to estimate a local treatment effect at a cutoff (e.g.~income eligibility threshold) \cite{calonico_optimal_2020}, and can also be applied if the regression process is a probability density function \cite{cattaneo_simple_2020}.
Whereas these applications deal with estimating a function at a single point, our work analyzes the downstream effects of using LPR as a filter to recover a continuous-time signal at all points simultaneously.
% This

Another type of derivative approximation worth mentioning is the Modulating Function method, which uses integration by parts to pass derivatives onto a test function \cite{unbehauen_identification_1997,niethammer_parameter_2001}.
This manipulation amounts to computing the (Schwarz) weak derivative of the interpolated noisy data.

\section{Proof of Lemma~\ref{lem:consistent-sum}}
% \begin{pf}
  Let \(\Delta \tilde x_i = \hat x_i - \expect \hat x_i\) and \(\bar x_i = \expect \hat x_i\).
  For contracting tensors, we use implicit summation with abstract index notation, where the variable \(\mu_1\) ranges over \([0\ldots m]\).
  By Taylor expansion,
  \begin{align}
    \begin{split}
      \MoveEqLeft
      \frac{1}{n'} \sum_{i = 1}^{n'} f(\hat x_i)
      \\
       & =
      \underbrace{
        \frac{1}{n'} \sum_{i = 1}^{n'} f(\bar x_i)
      }_{\text{I}}
      % \\
      % &\quad 
      + \underbrace{
      \frac{1}{n'} \sum_{i = 1}^{n'}
      \partial_{\mu_1} f(\bar x_i) \Delta \tilde x_i^{\mu_1}
      }_{
      \text{II}
      }
      \\
       & \quad
      + \underbrace{
        \frac{1}{n'} \sum_{i = 1}^{n'}
        O(\left\|\Delta \tilde x_i\right\|^2)
      }_{\text{III}}
      \\
    \end{split}
    \label{eq:consistent-sum}
  \end{align}
  Using the bias hypothesis of consistent filtering, \(f(\bar x_i) = f(x_i) + O(h^\beta)\).
  This settles the first sum (I).
  The third term (III) is \(O_{p}(h^{2\gamma})\) by the fluctuation hypothesis of consistent filtering.
  The sum (II) has \(n' \sim n\) with a local dependence structure: summands \(i\) and \(j\) are dependent if \(|i - j| \leq N \sim h^{-\alpha}\).
  We split the sum into \(N\) different sums of \(O(n/N)\) independent terms:
  \begin{align}
    \begin{split}
      E_6 := \MoveEqLeft\frac{1}{n'} \sum_{i = 1}^{n'}
      \partial_{\mu_1} f(\bar x_i) \Delta \tilde x_i^{\mu_1}
      \\
       & =
      \frac{1}{n'}
      \sum_{i = 1}^{N}
      \sum_{\ell = 0}^{n/N - 1}
      \underbrace{\partial_{\mu_1}  f(\bar x_{N\ell + i})}_{O(1)}
      \underbrace{\Delta \tilde x_{N\ell + i}^{\mu_1}}_{O_p(h^\gamma)}
      \label{eq:dependent-sum}
    \end{split}
    \intertext{By independence, the inner sum is \(O_p(h^\gamma(n/N)^{1/2})\). Using \(N \sim h^{-\alpha}\) and \(n' \sim n\),}
     E_6
     & = h^{-\alpha} n^{-1} O_p(h^\gamma (n/h^{-\alpha})^{1/2})
    \\
     & = O_p(n^{-1/2} h^{- \frac{1}{2}\alpha + \gamma}).
  \end{align}
  The conclusion follows from substituting (I), (II), and (III) into \eqref{eq:consistent-sum}.
% \end{pf}

\section{Proof of Lemma~\ref{lem:consistent-sum-2nd}}
For contracting tensors, we use implicit summation with abstract index notation, where the variables \(\mu_1, \mu_2\) range over \([0\ldots m]\).
  A quadratic Taylor expansion yields
  \begin{align}
    \begin{split}
      \MoveEqLeft
      \frac{1}{n'} \sum_{i = 1}^{n'} f(\hat x_i)
      \\
       & =
      \underbrace{
        \frac{1}{n'} \sum_{i = 1}^{n'} f(\bar x_i)
      }_{\text{I}}
      % \\
      % &\quad 
      + \underbrace{
      \frac{1}{n'} \sum_{i = 1}^{n'}
      \partial_{\mu_1} f(\bar x_i) \Delta \tilde x_i^{\mu_1}
      }_{
      \text{II}
      }
      \\
       & \quad
      + \underbrace{
      \frac{1}{2n'} \sum_{i = 1}^{n'}
      \partial_{\mu_1} \partial_{\mu_2} f(\bar x_i)
      \Delta \tilde x_i^{\mu_1}
      \Delta \tilde x_i^{\mu_2}
      }_{
      \text{III}
      }
      \\
       & \quad
      + \underbrace{
        \frac{1}{n'} \sum_{i = 1}^{n'}
        O(\left\|\Delta \tilde x_i\right\|^3)
      }_{\text{IV}}
      \\
    \end{split}
    \label{eq:consistent-sum-2nd}
  \end{align}
  % The terms (I) and (II) obey the same bounds in Lem.~\ref{lem:consistent-sum}.
  As in Lem.~\ref{lem:consistent-sum}, the terms (I) and (II) are \(O(h^\beta)\) and \(O_p(n^{-1/2}h^{-\frac{1}{2}\alpha + \gamma})\), respectively.
  Fact~\ref{fact:dependent-sum} shows that the fourth moment hypothesis of consistent filtering implies term (III) deviates from its expectation by \(O_p(n^{-1/2}h^{-\frac{1}{2}\alpha + 2\gamma})\).
  Finally, the term (IV) is bounded as in Lem.~\ref{lem:consistent-sum},
  resulting in
  \begin{multline}
    \frac{1}{n'} \sum_{i = 1}^{n'} f(\hat x_i)
    - \frac{1}{n'} \sum_{i = 1}^{n'} f(x_i)
    \\
    = \frac{1}{2n'} \sum_{i = 1}^{n'} \partial_{\mu_1}\partial_{\mu_2} f(\bar  x_i) \Sigma^{\mu_1\mu_2}
    \\
    + O_p(h^\beta + n^{-1/2} h^{- \frac{1}{2}\alpha + \gamma} + h^{3\gamma}).
    \label{eq:sum-2nd}
  \end{multline}

  Applying Lem.~\ref{lem:consistent-sum} to the function \(\partial_{\mu_1}\partial_{\mu_2} f \) yields
  \begin{align}
    \begin{split}
      \MoveEqLeft
      \frac{1}{n'} \sum_{i = 1}^{n'}
      \partial_{\mu_1}\partial_{\mu_2} f(\hat x_i)
      - \frac{1}{n'} \sum_{i = 1}^{n'}
      \partial_{\mu_1}\partial_{\mu_2} f(\bar x_i)
      \\
       & = O_p(h^\beta + n^{-1/2} h^{- \frac{1}{2}\alpha + \gamma} + h^{3\gamma})
    \end{split}
    \intertext{Multiplying both sides by \(\frac{\Sigma^{\mu_1\mu_2}}{2} = O(h^{2\gamma})\),}
    \begin{split}
      \MoveEqLeft
      \frac{1}{n'} \sum_{i = 1}^{n'}
      \frac{\Sigma^{\mu_1\mu_2}}{2}
      \partial_{\mu_1}\partial_{\mu_2} f(\hat x_i)
      - \frac{1}{n'} \sum_{i = 1}^{n'}
      \frac{\Sigma^{\mu_1\mu_2}}{2}
      \partial_{\mu_1}\partial_{\mu_2} f(\bar x_i)
      \\
       & = h^{2\gamma} O_p(h^\beta + n^{-1/2} h^{- \frac{1}{2}\alpha + \gamma} + h^{3\gamma}).
    \end{split}
    \label{eq:sum-bias-2nd}
  \end{align}
  This shows that the error in the second-order term is stochastically negligible.

  To get the conclusion, subtract this equation from \eqref{eq:sum-2nd}.
\section{Filtering details}
The construction in this section are provided for thoroughness and extend the ideas in \cite{brabanter_derivative_nodate}.
% , but essentially mirror the ideas in \cite{brabanter_derivative_nodate} and other local polynomial formulas for differentiation with smoothing.
In the signal processing point of view, local polynomial regression is carried out on a fixed grid, affording easier analysis than random designs in statistics (e.g.~\cite{fan_local_2003}).
% It is similar to local polynomial regression as used in statistics (e.g.~\cite{fan_local_2003}) because the independent variable is sampled on a fixed grid, rather than according to an unknown probability density function.

\label{section:filtering-details}

\label{section:accurate-time}
For each \(d \in [0\ldots m]\), the \(N\) coefficients \(D^d_{k}\), \(k \in [1\ldots N]\), may be selected to solve up to \(N\) independent equations.
In order for \(D\) to attain the desired \(p\)th order accuracy, it must correctly differentiate polynomials of degrees \([0\ldots p - 1]\).
Specifically, we require that given values at times \((n_i + 1) h, (n_i + 1) h, \ldots, (n_i + N) h\), the filter yields a derivative estimate at time \(t_i = i_0 h/2\) for a desired position \(i_0\), such as \(i_0 = (N + 1)/2\).

% \begin{rem}
%   With \(y\) has sufficient regularity, the odd structure of a centered Taylor expansion provides an extra degree of accuracy ``for free.'' \cite{fan_local_2003}
%   We leave this out of the analysis for simplicity.
% \end{rem}

These \(p\) linear constraints can be called ``natural conditions'' \cite{soderstrom_least_1997}.
With the remaining degrees of freedom, we minimize some convex matrix norm \(f(D)\).
In our work, we minimize the Frobenius norm.
We reserve possibilities such as the general operator norm (induced by any two norms on \(\mathbb{R}^{N}\) and \(\mathbb{R}^{m + 1}\)) and Schatten norms for future work, as well as the option to
impose input and output weights by \(f(W_\text{out}DW_\text{in})\), perhaps tuned from data.

We prescribe \(D\) as a solution to the following convex program:
\begin{align}
  \label{eq:D-program}
  \begin{split}
     & \min_{D \in \mathbb{R}^{m \times N}} \quad f(D)
    \\
     & \operatorname{subject\ to}\quad
    DA = B
  \end{split}
\end{align}
where \(A \in \mathbb{R}^{N \times p}\) and \(B \in \mathbb{R}^{(m+1) \times p}\) are given by
\begin{subequations}
  \label{eq:D-program-constraints}
  \begin{align}
    A_{ij}
     & =
    (i - i_0)^j h^j / j!
    \\
    B^d_{j}
     & =
    \delta_{dj}
    \\
    i
     & \in [1\ldots N]
    \\
    j
     & \in [0\ldots p - 1]
    \\
    d
     & \in [0\ldots m]
  \end{align}
\end{subequations}

% \begin{rem}
%   As Lemma~\ref{lemma:D-formula-bound} shows, \(D\) can be viewed as local polynomial regression with a boxcar window function.
%   While this formula is not novel, we have arrived at it via a different path.
%   In local polynomial regression, a polynomial fit is used to minimize the sum of squared residual.
%   In our problem, \(D\) is chosen to satisfy a number of exact conditions (bias) and otherwise have minimum norm (variance).
%   It is interesting and not entirely \emph{a priori} obvious that these two paths lead to the same destination.
%   % Point out that frob minimum is an op minimizer. bias depends on op, variance on frob
% \end{rem}

\begin{rem}[Numerics of \(D\)]
  Even though Lemma~\ref{lemma:D-formula-bound} gives an explicit formula for a certain version of \(D\),
   \(A^\intercal A\) contains the infamously ill-conditioned Hilbert matrix \eqref{eq:hilbert-matrix}.
  %  it is more numerically stable in practice to solve \eqref{eq:D-program} using a division-free convex optimization utility.
  For numerical stability, we solve for \(D\) by rewriting the natural conditions \eqref{eq:D-program-constraints} in a basis of Legendre polynomials.
  % , but we have found this further caution unnecessary for values of \(p\) up to \(10\), which is more than enough for most applications.
\end{rem}

% In the following analysis and in applications, \(D\), unless otherwise specified, will be solved for \(f(D) = \left\|D\right\|\), the induced (operator) 2-norm.\footnote{
%   \color{red} We go through the motions of analyzing what is essentially a Savitzky-Golay filter that simultaneously outputs derivatives along with the smoothed signal.
% A bias-variance analysis for differentiation is used in recent literature \cite{krishnan_selection_2013,john_adaptive_2021}.
% Some similar results are in \cite[Theorem~3.1]{fan_local_2003}, but they do not account for Sobolev regularity (Assumption~\ref{assum:sobolev}) or arbitrary \(i_0\) leading to a non-symmetric filter. This book uses exact weighting kernels which are less optimal than our convex program, as well as a more more simplistic automatic tuning.
% The convex programming filter specification is more universal and arguably lends itself to more insightful proofs.
% }
% This is the quantity of actual interest, used in Lemmas~\ref{lem:acc-diff-bias} and \ref{lemma:acc-diff-variance}, but our indirect approach is to majorize the operator norm with the Frobenius norm.

\begin{lem}[Row-by-row bound on \(D\)]
  \label{lemma:D-formula-bound}
  The solution to \eqref{eq:D-program} using \(f(D) = \left\|D\right\|_{\frob}\) is
  \begin{align*}
    D & = B (A^\intercal A)^{-1} A^\intercal
    \intertext{and satisfies}
    \left\|e_d^\intercal D\right\|
      & \leq C(m, p) N^{-d - \frac{1}{2}} h^{-d}
    \intertext{and}
    \left\|D\right\|
      & \leq C(m, p) N^{-m - \frac{1}{2}} h^{-m}.
  \end{align*}
\end{lem}
\begin{pf}
  \textbf{Formula for \(D\)}
  Write the Frobenius inner product as \(\left\langle X, Y\right\rangle_{\frob} = \trace \del{X^\intercal Y}\).
  Let \(\Lambda \in \mathbb{R}^{(m + 1)\times p}\) be a Lagrange multiplier, and
  form the Lagrangian \(\frac{1}{2} \left\langle D, D\right\rangle_{\frob} - \left\langle \Lambda, DA - B\right\rangle_{\frob}\).
  First-order optimality yields \(D = \Lambda A^\intercal\).
  Right-multiplying by \(A\), we get \(B = \Lambda (A^\intercal A)\) which can be solved for \(\Lambda\).

  \textbf{Estimating involving \(\tilde A\) and \(\tilde B\)}
  Rescale \eqref{eq:D-program-constraints} in order to normalize \(A^\intercal A\):
  \begin{subequations}
    \begin{align}
      \tilde A_{ij}
       & =
      (i - i_0)^j N^{-j}
      \\
      \tilde B^d_{j}
       & =
      \delta_{dj} N^{-j} h^{-j} d!
    \end{align}
  \end{subequations}

  To estimate \(\tilde A^\intercal \tilde A\), notice that
  \begin{align}
    \del{\tilde A^\intercal \tilde A}_{jk}
     & = \sum_{i = 1}^N \del{\frac{i - i_0}{N}}^{j + k}
    \\
    \intertext{is a right Riemann sum. Evaluating the integral (with an error estimate),}
        \del{\tilde A^\intercal \tilde A}_{jk}
     & = \frac{N}{j + k + 1} + S_{jk}                   \\
    \left|S_{jk}\right|
     & \leq \frac{2p}{N}.
    \label{eq:hilbert-matrix}
    \intertext{As a consequence of this rescaling, we have the estimate}
    \left\| \del{\tilde A^\intercal \tilde A}^{-1} \right\|
     & \leq C(p) N^{-1}.
    \label{eq:tilde-A-bound}
  \end{align}
  % where the constant \(C(p)\) is technically computable in finitely many operations by virtue of the fact that Weyl's inequality controls pertubations of the spectrum of \(\tilde A^\intercal \tilde A\), allowing for a supremum over \(N\).

  % To estimate \(\tilde B\), note that this matrix is diagonal save for a block of zeros.
  % Taking the highest power of \(N\) and the lowest power of \(h\),
  % \begin{align}
  %   \left\| \tilde B\right\|
  %    & \leq
  %   m! h^{-m}.
  %   \label{eq:tilde-B-bound}
  % \end{align}

  \textbf{Bounding \(D\)}
  Starting from \(D = B(A^\intercal A)^{-1} A^\intercal\),
  \begin{align}
    \left\|e_d^\intercal D\right\|
     & = \sqrt{\left\|e_d^\intercal B(A^\intercal A)^{-1} B^\intercal e_d\right\| }
    \\
     & \leq \left\|e_d^\intercal B\right\| \left\|(A^\intercal A)^{-1}\right\|
  \end{align}
  The conclusion follows after using \eqref{eq:tilde-A-bound} and the following fact: because \(\tilde B\) is diagonal (except for a block of zeros), we have \(\left\|e_d^\intercal \tilde B\right\| = N^{-d} h^{-d} d!\).

  % \begin{align}
  %   \left\|D\right\|^2_{\frob}
  %    & = \trace \tilde B (\tilde A^\intercal \tilde A)^{-1} \tilde B^\intercal
  %   \\
  %    & = \trace \tilde B^\intercal \tilde B (\tilde A^\intercal \tilde A)^{-1}
  %   \\
  %   \intertext{By the Cauchy-Schwarz inequality,}
  %    & \leq \sqrt{\trace \del{\tilde B^\intercal \tilde B}^2 \trace (\tilde A^\intercal \tilde A)^{-2}}
  %   \\
  % \end{align}
\end{pf}

% \begin{rem}
%   In the following Lemma, the regularity of \(y\) enters as the quantity \(p - 1/q\), similar to the one-dimensional case of Morrey's inequality.
% \end{rem}

\begin{lem}[Bias]
  \label{lem:acc-diff-bias}
  Let \(\hat x_j\) be defined using \(D\), the solution of \eqref{eq:D-program}. Then
  \begin{align*}
    \left|\expect \hat x_j - x_j\right|
     & \leq
    C(m, p) R_p (Nh)^{p - m}
    \\
     & =
    C(m, p) R_p h^{(p - m)(1 - \alpha)}.
  \end{align*}
\end{lem}
\begin{pf}
  This proof resembles  that of Lemma~\ref{lemma:diff-bias}, with the main difference that no matter the \(d\), \(x^d_j\) will admit a \((p-1)\)th degree Taylor expansion.
  Around \(t_j = (n_j + i_0)h\),
  \begin{align}
    \expect z_{n_j + k}
     & = \sum_{\nu = 0}^{p - 1}
    \frac{y^{(\nu)}(t_j)}{\nu!}((k - i_0)h)^\nu + R(k),
    \\
    \intertext{where the remainder obeys the estimate}
    R(k)
     & \leq
    R_{p} C(m, p, q) (Nh)^{p}.
    \label{eq:D-remainder}
    \intertext{Contracting the \(d\)th row of \(D\) with \(z_{[n_i + 1\ldots n_i+N]}\),}
    \begin{split}
      \expect \hat x^d_j
       & = \sum_{j = 1}^N D^d_k
      \sum_{\nu = 0}^{p - 1}
      \frac{y^{(\nu)}(t_j)}{\nu!}((k - i_0)h)^\nu
      \\
       & \quad
      +\sum_{k = 1}^N D^d_k  R(k)
    \end{split}
    \intertext{The natural conditions \eqref{eq:D-program-constraints} ensure that}
    \expect \hat x^d_j
     & = x^d(t_j) + \sum_{k = 1}^N D^d_k  R(k)
    \\
    \intertext{Applying Lemma~\ref{lemma:D-formula-bound},}
    \left|\expect \Delta x_i^d\right|
     & \leq C(m, p, q) R_p  (Nh)^{p - m }.
  \end{align}
  Finally, the conclusion follows from adding up \(d \in [0\ldots m]\) and applying \(N \sim h^{-\alpha}\).
\end{pf}
% {\color{red} should the power be more negative in \(N\)? use the \(n/N\) trick}
\begin{lem}[Fluctuation]
  \label{lemma:acc-diff-variance}
  Let \(\hat x_j\) be defined using \(D\), the solution of \eqref{eq:D-program}. Then
  Then the filtered estimate satisfies
  \begin{align*}
    \expect \left\|\hat x_j - \expect \hat x_j\right\|^4
     & \leq  \sigma^4 N^{-4m - 2} h^{-4m}
    \\
     & =  \sigma^4 h^{(4m + 2)\alpha} h^{-4m}
  \end{align*}
\end{lem}
\begin{pf}
  Arrange this window's noise terms in a random vector according to \(W_j = w_{(n_j + k) h}\).
  Using Lemma~\ref{lemma:D-formula-bound},
  \begin{align}
    \expect \left\|\hat x_j - \expect \hat x_j\right\|^4
     & =
    \expect \left\|D W\right\|^4
    \\
     & \leq C(m) \sigma^4 n^2 \left\|D\right\|^4
  \end{align}
  The conclusion follows after applying Lemma~\ref{lemma:D-formula-bound} for \(\left\|D\right\|\).
\end{pf}

\bibliographystyle{plain}        % Include this if you use bibtex 
\bibliography{export}           % and a bib file to produce the 
                                 % bibliography (preferred). The
                                 % correct style is generated by
                                 % Elsevier at the time of printing.

%\begin{thebibliography}{99}     % Otherwise use the  
                                 % thebibliography environment.
                                 % Insert the full references here.
                                 % See a recent issue of Automatica 
                                 % for the style.
%  \bibitem[Heritage, 1992]{Heritage:92}
%     (1992) {\it The American Heritage. 
%     Dictionary of the American Language.}
%     Houghton Mifflin Company.
%  \bibitem[Able, 1956]{Abl:56}
%     B.~C.~Able (1956). Nucleic acid content of macroscope. 
%     {\it Nature 2}, 7--9. 
%  \bibitem[Able {\em et al.}, 1954]{AbTaRu:54}   
%     B.~C. Able, R.~A. Tagg, and M.~Rush (1954).
%     Enzyme-catalyzed cellular transanimations.
%     In A.~F.~Round, editor, 
%     {\it Advances in Enzymology Vol. 2} (125--247). 
%     New York, Academic Press.
%  \bibitem[R.~Keohane, 1958]{Keo:58}
%     R.~Keohane (1958).
%     {\it Power and Interdependence: 
%     World Politics in Transition.}
%     Boston, Little, Brown \& Co.
%  \bibitem[Powers, 1985]{Pow:85}
%     T.~Powers (1985).
%     Is there a way out?
%     {\it Harpers, June 1985}, 35--47.

%\end{thebibliography}

% \appendix                      % in the appendices.
\end{document}